\documentstyle[prl,aps,preprint]{revtex}
\begin{document}

\draft 
\tightenlines
\title{Self-dual gravity and self-dual Yang-Mills\\ in the context 
of Macdowell-Mansouri formalism }
\author{ J.A. Nieto$^\star$\thanks{E-Mail: nieto@uas.uasnet.mx} 
and J. Socorro$^{\$}$\thanks{E-mail: socorro@ifug4.ugto.mx}\\
$^{\star}$ Facultad de Ciencias F\'{\i}sico-Matem\'{a}ticas \\
 de la Universidad Aut\'{o}noma de Sinaloa, C.P. 80010, Culiac\'{a}n, 
Sinaloa, M\'{e}xico.\\
$^{\$}$ Instituto de F\'{\i}sica de la Universidad de Guanajuato,\\
Apartado Postal E-143, C.P. 37150, Le\'on, Guanajuato, Mexico.
}
\date{\today}

\maketitle

\widetext
\begin{abstract}

In this work we propose an action which unifies self-dual gravity and
self-dual Yang-Mills in the context of the Macdowell-Mansouri 
formalism. We claim that such an action may be used to find the S-dual 
action for both self-dual gravity and self-dual Yang-Mills.
\end{abstract}

\pacs{Pacs No.: 04.50.+h, 04.90.+e \hspace{7cm}}

\narrowtext

 Recently, using a self-dual generalization of the 
Mac\-do\-well\--Man\-sou\-ri
formalism \cite{NOS1,NOS2} gravitational S-duality has been considered by
Garc\'{\i}a-Compean, Obreg\'on, Plebanski and Ram\'{\i}rez (GOPR)
\cite{GC1,GC2}. The central
idea of these authors was to find a dual gravitational model by using a
similar mechanism to the one used in the case of self-dual Yang-Mills theory
\cite{Q}. From a fundamental point of view, however, it will be more 
satisfactory to have S-dual invariant unification model for gravity and 
Yang-Mills. As a first step in this direction we may attempt to have
unified theory for self-dual gravity and self-dual Yang-Mills in the context
of Macdowell-Mansouri formalism \cite{MaMa,Ma} 
(see also \cite{To,We,Cha,Pa,KiSt,Fr,GH,Ha,ObHe,Soo} ). But, to the
best of our knowledge, such an unified theory is lacking in the literature.
In this paper we propose an action in the Macdowell-Mansouri context which
not only reduces to self-dual gravitational action and self-dual Yang-Mills
but also gives topological terms and interesting interacting terms between
gravitational curvature and Yang-Mills field strength.

Let us first recall some aspects of the Macdowell-Mansouri's gravitational
formalism. It is known that such a formalism mimics the Yang-Mills-type
gauge theory in four space-time dimensions as much as possible and that it
has been successfully applied to construct different supergravity theories
\cite{Ni}. The central idea is to represent the gravitational field as a
connection one-form associated to some group that contains the Lorentz group
as a subgroup. The typical example is provided by the SO(3,2) anti-de Sitter
gauge theory of gravity. In this case the SO(3,2) gravitational gauge field 
$\omega _\mu ^{AB}=-$ $\omega _\mu ^{BA}$ is broken into the SO(3,1)
connection $\omega _\mu ^{ab}$  and the 
$\omega _\mu ^{4a}=e_\mu ^a$ vierbein
field. Thus, the anti-de Sitter cur\-va\-tu\-re 
\begin{equation}
{\cal R}_{\mu \nu }^{AB}=\partial_\mu \omega_\nu ^{AB}- 
\partial_\nu \omega_\mu^{AB} + \omega_\mu^{AC} 
\omega_{\nu C}\,^{ B} - \omega_\mu^{BC} \omega_{\nu C}\,^A  
\label{ec.1}
\end{equation}
leads to

\begin{equation}
{\cal R}_{\mu \nu }^{ab}=R_{\mu \nu }^{ab}+\Sigma _{\mu \nu }^{ab}
\label{ec.2}
\end{equation}
and

\begin{equation}
{\cal R}_{\mu \nu }^{4a}=\partial _\mu e_\nu ^a-\partial _\nu e_\mu
^a+\omega _\mu ^{ac}e_{\nu c}^{}-\omega _\nu ^{ac}e_{\mu c}^{},  \label{ec.3}
\end{equation}
where

\begin{equation}
R_{\mu \nu }^{ab}=\partial _\mu \omega _\nu ^{ab}-\partial _\nu 
\omega_\mu^{ab}+\omega_\mu^{ac}\omega_{\nu c}\,^b-\omega_\mu^{bc}
 \omega_{\nu c}\,^a  
\label{ec.4}
\end{equation}
is the SO(3,1) curvature and

\begin{equation}
\Sigma _{\mu \nu }^{ab}=e_\mu ^a \, e_\nu ^b - e_\mu ^b \, e_\nu ^a \, .  
\label{ec.5}
\end{equation}
It turns out that ${\cal R}_{\mu \nu }^{4a}=T_{\mu \nu }^a$ can be
identified with the torsion.

The Macdowell-Mansouri's action is

\begin{equation}
S=\frac 14 \int_M \varepsilon^{\mu \nu \alpha \beta }\, 
{\cal R}_{\mu \nu }^{ab} \, {\cal R}_{\alpha \beta }^{cd}\epsilon_{abcd},
 \label{ec.6}
\end{equation}
where $\varepsilon ^{\mu \nu \alpha \beta }$ is the fully antisymmetric
tensor associated to the space-time, with $\varepsilon ^{0123}=1,$ while 
$\epsilon _{abcd}$ is also the fully antisymmetric tensor but now associated
to the internal group SO(3,1), with $\epsilon _{0123}=-1.$ We assume that the
internal metric is given by $\eta _{ab}=(-1,1,1,1).$ Therefore, we have 
$\epsilon ^{0123}=1.$ It is well known that, using (\ref{ec.2}-\ref{ec.5}), 
the action (\ref{ec.6})
leads to three terms; the Hilbert-Einstein action, the cosmological constant
term and the Euler topological invariant (or Gauss-Bonnet term). It is worth
mentioning that the action (\ref{ec.6}) may also be obtained u\-sing 
Lovelock theory
(see \cite{TeZa} and references there in).

The action (\ref{ec.6}) has been generalized in the form \cite{NOS1}

\begin{equation}
S=\frac 14 \int_M \varepsilon^{\mu \nu \alpha \beta }
{^+{\cal R}}_{\mu \nu }^{ab}\, {^+{\cal R}}_{\alpha \beta}^{cd}
\epsilon_{abcd} \, .  
\label{ec.7}
\end{equation}
Here, $^+{\cal R}_{\mu \nu }^{ab}$ is given by

\begin{equation}
^{\pm }{\cal R}_{\mu \nu }^{ab} =\frac{1}{2}{^\pm M}_{cd}^{ab}\,
{\cal R}_{\alpha \beta }^{cd}\, ,  
\label{ec.8}
\end{equation}
where

\begin{equation}
^{\pm }M_{cd}^{ab}=\delta_{cd}^{ab}\mp i\epsilon^{ab}\, _{cd},
\label{ec.9}
\end{equation}
with $\delta _{cd}^{ab}=\delta _c^a\delta _d^b-\delta _c^b\delta _d^a.$ It
turns out that $^{+}{\cal R}_{\mu \nu }^{ab}$ is self-dual, while 
$^{-}{\cal R}_{\mu \nu }^{ab}$ is anti self-dual curvature. 
Therefore, the action (\ref{ec.7})
describes self-dual gravity. In fact, in reference \cite{NOS2}, 
it was shown that the action (\ref{ec.7}) can be decomposed in four 
terms: the Euler topological
invariant, the Pontrjagin topological invariant, the Ashtekar action (as
proposed by Samuel \cite{Sa} and Jacobson and Smolin \cite{JaSm}) and the usual
cosmological constant term.

In order to develope a S-dual gravitational action, the key idea in
reference \cite{GC1} was to consider the generalized action

\begin{equation}
S=\frac 14{^+\tau} \int_M \varepsilon^{\mu \nu \alpha \beta }\,
{^+{\cal R}}_{\mu \nu }^{ab}\,{^+{\cal R}}_{\alpha \beta }^{cd}
\epsilon_{abcd} 
-\frac 14{^-\tau} \int_M \varepsilon^{\mu \nu \alpha \beta }\,
{^-{\cal R}}_{\mu \nu }^{ab}\, {^-{\cal R}}_{\alpha \beta}^{cd}
\epsilon_{abcd}\, ,
 \label{ec.10}
\end{equation}
where $^{+}\tau $ and $^{-}\tau $ are two different constant parameters. In
fact, let us briefly mention how GOPR found the S-dual 
gravitational model. First, consider the identity

\begin{equation}
\frac 14 {^\pm M}_{ef}^{ab}\, {^\pm M}_{gh}^{cd}\epsilon_{abcd}
=2(\epsilon_{efgh}\pm i\delta_{ef,gh})\, ,  
\label{ec.11}
\end{equation}
where $\delta _{ef,gh}=\eta _{eg}\eta _{fh}-\eta _{eh}\eta _{fg}.$ It is not
difficult to see that (\ref{ec.10}) can be rewritten as

\begin{equation}
S=\frac 12(^{+}\tau -^{-}\tau )\int_M \varepsilon^{\mu \nu \alpha \beta }
\, {\cal R}_{\mu \nu}^{ab}\, {\cal R}_{\alpha \beta}^{cd}
\epsilon_{abcd} 
-\frac i2(^{+}\tau +^{-}\tau )\int_M
\varepsilon^{\mu \nu \alpha \beta }\, {\cal R}_{\mu \nu }^{ab}\,
{^{*}{\cal R}}_{\alpha \beta }^{cd}\, \epsilon _{abcd},
\label{ec.12}
\end{equation}
where

\begin{equation}
^{\ast }{\cal R}_{\alpha \beta }^{ab}=\frac 12
\epsilon^{ab}\,_{cd} \,{\cal R}_{\alpha \beta}^{cd}.  
\label{ec.13}
\end{equation}
The second term in (\ref{ec.11}) can be identified with a 
$\theta $ term, with $\theta $ given by $\frac 12(^{+}\tau +^{-}\tau ).$ 
We may follow then a
similar procedure as in the case of self-dual Yang-Mills 
\cite{Wi,GanSo,Mo}. In our
case, such a procedure consists in writing first the Palatini type action

\begin{equation}
S=\int_M \varepsilon^{\mu \nu \alpha \beta}\left(
c_1\, ^{+}G_{\mu \nu}^{ab}\, {^+G}_{\alpha \beta}^{cd} + c_2\,
{^-G}_{\mu \nu }^{ab}\,{^-G}_{\alpha \beta}^{cd}
 + c_3\,^{+}{\cal R}_{\mu \nu }^{ab}\,
{^+G}_{\alpha \beta }^{cd}+c_4\, {^-{\cal R}}_{\mu \nu }^{ab}\, 
{^-G}_{\alpha \beta }^{cd})\epsilon_{abcd}\right),
\label{ec.14}
\end{equation}
where c$_1,$c$_2,$c$_3$ and c$_4$ are constants, and then considering the
partition function

\begin{equation}
Z=\int d{^+G}\,\,d{^-G}\, d\omega \,e^{-S}.  
\label{ec.15}
\end{equation}
where the action $S$ is given by (\ref{ec.14}). Thus, after performing a 
partial integration over $\omega $ we obtain the desired dual action, which 
relevant part is given by

\begin{equation}
S=\int_M \varepsilon^{\mu \nu \alpha \beta }\left(
-\frac{1}{4^+\tau}\, {^+G}_{\mu \nu}^{ab}\,\,{^+G}_{\alpha \beta}^{cd}
 +\frac{1}{4^-\tau} \,{^-G}_{\mu \nu}^{ab}\,\,
{^-G}_{\alpha \beta }^{cd}+\cdots \right)\, \epsilon_{abcd}\, ,  
\label{ec.16}
\end{equation}
(see reference \cite{GC1} for details). (It is worth mentioning that the action
(\ref{ec.10}) can be recovered from the action (\ref{ec.14}) by applying the 
Palatini procedure to the variables $^{+}$G and $^{-}$G.)

In order to unify self-dual Yang-Mill fields and self-dual gravity, we shall
generalize the action (\ref{ec.10}). Our proposed action is

\begin{equation}
S=\frac 14{^+\tau} \int_M E^{\mu \nu \alpha \beta }\,\,
{^+{\cal R}}_{\mu \nu}^{ab}\,\,{^+{\cal R}}_{\alpha \beta }^{cd}\,
\epsilon_{abcd}
 -\frac{1}{4}{^-\tau}\int_M E^{\mu \nu \alpha \beta}\,\,
{^-{\cal R}}_{\mu \nu }^{ab} \,\,{^-{\cal R}}_{\alpha \beta }^{cd}\,
\epsilon_{abcd},  
\label{ec.17}
\end{equation}
where

\begin{equation}
E^{\mu \nu \alpha \beta }=\varepsilon^{\mu \nu \alpha \beta } + 
ieF^{\mu \nu i} F^{\alpha \beta j} g_{ij}.  
\label{ec.18}
\end{equation}
with $e=\det (e_\mu ^a).$ Here, 
$F_{\mu \nu }^j=F^{\alpha \beta j}e_\mu ^ae_\alpha ^c e_\nu ^b 
e_\beta ^d\eta _{ac}\eta _{bd}$ is the field strength
which is given in terms of the gauge potential $A_\mu ^i$ as

\begin{equation}
F_{\mu \nu }^i=\partial _\mu A_\nu ^i-\partial _\nu A_\mu ^i+f_{kj}^iA_\mu
^k A_\nu ^j,  
\label{ec.19}
\end{equation}
where $f_{kj}^i$ are the structure constants of the Lie algebra of some
compact group G and $g_{ij}$ is the Cartan-Killing metric associated to G.

We would like to prove that the action (\ref{ec.17}) unifies self-dual 
gravity and self-dual Yang-Mills. For simplicity, let us first 
focus in the action

\begin{equation}
^{+}S=\frac{1}{4}{^+\tau} \int_M E^{\mu \nu \alpha \beta}\,
{^+{\cal R}}_{\mu \nu}^{ab}\,{^+{\cal R}}_{\alpha \beta }^{cd}\,
\epsilon _{abcd}\, .
\label{ec.20}
\end{equation}

Substituting (\ref{ec.18}) into (\ref{ec.20}) we find that

\begin{equation}
^{+}S=^{+}S_1+^{+}S_2\, ,  \label{ec.21}
\end{equation}
where

\begin{equation}
^{+}S_1=\frac{1}{4}{^+\tau} \int_M \varepsilon^{\mu \nu \alpha \beta}\,
{^+{\cal R}}_{\mu \nu }^{ab}\,{^+{\cal R}}_{\alpha \beta }^{cd}
\epsilon_{abcd} \, ,
\label{ec.22}
\end{equation}
and

\begin{equation}
^{+}S_2=\frac{i}{4}{^+\tau} \int_M e F^{\mu \nu i} F^{\alpha \beta j} 
g_{ij} \,\,{^+{\cal R}}_{\mu \nu }^{ab}\,\,{^+{\cal R}}_{\alpha \beta}^{cd}
\, \epsilon_{abcd} \, .  \label{ec.23}
\end{equation}
We notice that $^{+}S_1$ is just the first term of the action (\ref{ec.10}), 
which we already know leads to four terms: the Euler topological invariant, the
Pontrjagin topological invariant, the Ashtekar action (as proposed by Samuel
and Jacobson and Smolin) and the usual cosmo\-lo\-gi\-cal cons\-tant term. So, new
features of our proposed action (\ref{ec.17}) should emerge 
from $^{+}S_2.$ In order to discover them, let us first observe that 
using (\ref{ec.11}), $^{+}S_2$ becomes

\begin{equation}
^{+}S_2=\frac{i}{2}{^+\tau} \int_M e F^{\mu \nu i} F^{\alpha \beta j}
g_{ab}\,{\cal R}_{\mu \nu }^{ab}\,{\cal R}_{\alpha \beta }^{cd}\,
\epsilon_{abcd}
-\frac{1}{2}{^+\tau} \int_M e F^{\mu \nu i} F^{\alpha \beta j}
g_{ij}\,{\cal R}_{\mu\nu }^{ab} \, {\cal R}_{\alpha \beta }^{cd}
\delta_{ab,cd} \, .  
\label{ec.24}
\end{equation}
Now, using (\ref{ec.2}) we find

\begin{eqnarray}
^{+}S_2&=& \frac{1}{2} {^+\tau} \left[ i\int_M e F^{\mu \nu i} 
F^{\alpha \beta j}
g_{ij}\,(R_{\mu \nu }^{ab}+\Sigma_{\mu \nu }^{ab})
(R_{\alpha \beta}^{cd}+\Sigma _{\alpha \beta }^{cd})\epsilon_{abcd} \right. 
\nonumber\\
&&\mbox{} \left. - \int_M e F^{\mu \nu i} F^{\alpha \beta j}
g_{ij}\, (R_{\mu\nu}^{ab}+\Sigma_{\mu \nu}^{ab})(R_{\alpha \beta}^{cd}
+\Sigma_{\alpha \beta}^{cd})\delta_{ab,cd}\right]\, ,
\label{ec.25}
\end{eqnarray}
This expression can be written as

\begin{equation}
^{+}S_2=^{+}{\cal S}_1+^{+}{\cal S}_2+^{+}{\cal S}_3  \label{ec.26}
\end{equation}
with

\begin{equation}
^{+}{\cal S}_1=\frac{i}{2}{^+\tau} \int_M e F^{\mu \nu i} 
F^{\alpha \beta j} g_{ij}\,R_{\mu \nu }^{ab}\,R_{\alpha \beta}^{cd}
\epsilon_{abcd}
-\frac{1}{2}{^+\tau} \int_M e F^{\mu \nu i} F^{\alpha \beta j}
g_{ij}\,R_{\mu \nu}^{ab}\, R_{\alpha \beta }^{cd}\delta_{ab,cd}\, ,  
\label{ec.27}
\end{equation}

\begin{equation}
^{+}{\cal S}_2=i{^+\tau} \int_M e F^{\mu \nu i} F^{\alpha \beta j}
g_{ij}\, R_{\mu \nu }^{ab}\, \Sigma_{\alpha \beta }^{cd}\,\epsilon_{abcd} 
-{^+\tau} \int_M e F^{\mu \nu i} F^{\alpha \beta j} g_{ij}\,
R_{\mu \nu }^{ab}\, \Sigma_{\alpha \beta }^{cd} \, \delta_{ab,cd}\, ,  
\label{ec.28}
\end{equation}
and

\begin{equation}
^{+}{\cal S}_3=\frac{i}{2}{^+\tau} \int_M e F^{\mu \nu i}
F^{\alpha \beta j} g_{ij}\, \Sigma_{\mu \nu}^{ab}\, 
\Sigma_{\alpha \beta }^{cd} \,\epsilon_{abcd} 
-\frac{1}{2}{^+\tau} \int_M e F^{\mu \nu i} F^{\alpha \beta j}
g_{ij}\, \Sigma_{\mu \nu }^{ab}\, \Sigma_{\alpha \beta }^{cd}\, 
\delta_{ab,cd}.  
\label{ec.29}
\end{equation}
In relation to the curvature $R_{\mu \nu }^{ab},$ notice that $^{+}{\cal S}_1
$ is quadratic, $^{+}{\cal S}_2$ is linear and $^{+}{\cal S}_3$ is
independent. As far as we know interacting terms between the curvature $%
R_{\mu \nu }^{ab}$ and the Yang-Mills field strength $F^{\mu \nu i}$ similar
to $^{+}{\cal S}_1$ and $^{+}{\cal S}_2$ have already been considered in the
literature\cite{Dru,LaMy}. Specifically, the second term in 
$^{+}{\cal S}_2$ has
been studied in connection with particles moving faster than light 
\cite{DaSho}. On
the other hand we recognize that the second term of $^{+}{\cal S}_3$ is the
typical action for Yang-Mills fields, while the first term is the second
Chern topological invariant. In fact, using (\ref{ec.5}) it is not difficult 
to see that $^{+}{\cal S}_3$ can be written as

\begin{equation}
^{+}{\cal S}_3=4i{^+\tau} \int_M e F^{\mu \nu i}\,{^\star F}_{\mu \nu}^j
\, g_{ij}
 -4{^+\tau}\int_M e F^{\mu \nu i}\, F_{\mu \nu }^j \, g_{ij} \, ,
\label{ec.30}
\end{equation}
where

\begin{equation}
{^\star F}_{\mu \nu }^i=\frac{1}{2} \epsilon_{\mu \nu \alpha \beta } 
F^{\alpha \beta i} \, .  
\label{ec.31}
\end{equation}
Here, $\epsilon_{\mu \nu \alpha \beta }=e_\mu^a e_\nu^b e_\alpha^c 
e_\beta^d \, \epsilon_{abcd}.$

Now, the action (\ref{ec.17}) can be written as $S=S_1+S_2,$ where $S_1$ 
is given by (\ref{ec.10}) and

\begin{equation}
S_2=\frac{i}{4}{^+\tau} \int_M e F^{\mu \nu i} \, F^{\alpha \beta j}
g_{ij}\,\,{^+{\cal R}}_{\mu \nu }^{ab}\,\,{^+{\cal R}}_{\alpha \beta }^{cd}
\, \epsilon_{abcd} 
-\frac{i}{4}{^-\tau} \int_M e F^{\mu \nu i} \, F^{\alpha \beta j}
g_{ij}\,\,{^-{\cal R}}_{\mu \nu }^{ab}\,\,{^-{\cal R}}_{\alpha \beta }^{cd}
\, \epsilon_{abcd}.  
\label{ec.32}
\end{equation}
Thus, thanks to the change of sign in the identity (\ref{ec.11}), 
depending if we use ${^+M}_{ef}^{ab}$ or ${^-M}_{ef}^{ab},$ we discover 
that (\ref{ec.32}) leads to the result

\begin{equation}
^{}S_2={\cal S}_1+{\cal S}_2+{\cal S}_3  \label{ec.33}
\end{equation}
where

\begin{eqnarray}
{\cal S}_1&=&\frac{i}{2}({^+\tau} -{^-\tau} ) \int_M e F^{\mu \nu i} 
F^{\alpha \beta j} g_{ij}\, R_{\mu \nu }^{ab}\, R_{\alpha \beta }^{cd}\,
\epsilon_{abcd} \nonumber\\
&&\mbox{}-\frac{1}{2}({^+\tau} +{^-\tau} ) \int_M e F^{\mu \nu i}
F^{\alpha \beta j} g_{ij}\, R_{\mu \nu }^{ab} \, R_{\alpha \beta }^{cd}
\delta_{ab,cd},
\label{ec.34}
\end{eqnarray}

\begin{eqnarray}
{\cal S}_2&=&i({^+\tau} -{^-\tau} ) \int_M e F^{\mu \nu i} 
F^{\alpha \beta j} g_{ij}\, R_{\mu \nu }^{ab}\, \Sigma_{\alpha \beta }^{cd}
\epsilon_{abcd} \nonumber\\
&&\mbox{} -({^+\tau} +{^-\tau} )\int_M e F^{\mu \nu i} F^{\alpha \beta j}
g_{ij}\, R_{\mu \nu }^{ab}\, \Sigma_{\alpha \beta }^{cd}\, \delta _{ab,cd},
\label{ec.35}
\end{eqnarray}
and

\begin{eqnarray}
{\cal S}_3&=&\frac{i}{2}({^+\tau} -{^-\tau} )\int_M e F^{\mu \nu i}
F^{\alpha \beta j} g_{ij}\, \Sigma_{\mu \nu }^{ab}\, 
\Sigma_{\alpha \beta }^{cd}\, \epsilon_{abcd} \nonumber\\ 
&&\mbox{} - \frac{1}{2}({^+\tau} +{^-\tau} )\int_M e F^{\mu \nu i}
F^{\alpha \beta j} g_{ij}\, \Sigma_{\mu \nu }^{ab}\, 
\Sigma_{\alpha \beta }^{cd}\, \delta_{ab,cd} \, .  
\label{ec.36}
\end{eqnarray}

Focusing our attention in ${\cal S}_3={^+{\cal S}}_3+{^-{\cal S}}_3$ we
learn that

\begin{equation}
{\cal S}_3=4i({^+\tau} -{^-\tau} )\int_M e F^{\mu \nu i}\,
{^\star F}_{\mu \nu}^j g_{ij}  
- 4({^+\tau} +{^-\tau} )\int_M e F^{\mu \nu i} F_{\mu \nu}^j
g_{ij} \, ,  
\label{ec.37}
\end{equation}
We recognize the second term as the Yang-Mills action with coupling constant 
$\frac{1}{g^2}$ given by $4({^+\tau} +{^-\tau} ),$ while the first term is
the second Chern class topological invariant\cite{NS} which can be identified 
with the $\theta $ term. So, ${\cal S}_3$ describes self-dual Yang-Mills gauge
fields. Notice the importance to consider the anti-self-dual part in the
action (\ref{ec.17}). Thanks to it the couplings between the first term and 
second term in (\ref{ec.37}) are different.

In this way, we have achieved a unification between self-dual gravity and 
self-dual Yang Mills in the context of Macdowell-Mansouri formalism. Since
S-duality has been studied either using self-dual Yang-Mills or self-dual
gravity our results suggest a more general S-duality for both self-dual
fields. Following to GOPR procedure\cite{GC1,GC2}, 
let us outline this idea.

To get the dual action of (\ref{ec.17}), we should first consider the 
Palatini type  action

\begin{equation}
S=\int_M E^{\mu \nu \alpha \beta }\left( c_1\, {^+G}_{\mu \nu }^{ab} \,
{^+G}_{\alpha \beta}^{cd}+ c_2\,{^-G}_{\mu \nu }^{ab}\,
{^-G}_{\alpha \beta }^{cd} 
 + c_3\,{^+{\cal R}}_{\mu \nu }^{ab}\,
{^+G}_{\alpha \beta }^{cd} + c_4\, {^-{\cal R}}_{\mu \nu}^{ab}\,
{^-G}_{\alpha \beta }^{cd} \right)\epsilon_{abcd}  \label{ec.38}
\end{equation}
and then to calculate the partition function

\begin{equation}
Z=\int d{^+G} \,\, d{^-G}\, dA\, d\omega e^{-S} \, ,
\label{ec.39}
\end{equation}
where the action $S$ is now given by (\ref{ec.38}). At present, we are 
investigating this possibility and we hope to report our results in the near 
future.

Let us conclude and make some final comments. In this work, we propose the
action (\ref{ec.17}). This action is reduced to the following terms: 
Ashtekar action, Euler Topological invariant, Pontrjagin topological 
invariant, cosmological
constant term, Yang-Mills action, second Chern class topological invariant,
and a number of interacting terms between the field strength 
$F_{\mu \nu }^i$
and the curvature ${\cal R}_{\mu \nu }^{ab}.$ In other words, with the
action (\ref{ec.17}) we have achieved a unified theory of self-dual gravity 
and self-dual Yang-Mills in the context of Macdowell-Mansouri formalism. From
this result, we expect to be able to study S-duality following the 
GOPR procedure\cite{GC1,GC2}.

\acknowledgments

We would like to thank to H. Uriarte and N. Alejo for helpful discussions.
This work was partially supported by CONACyT grant No. 3898P-E9608.

\end{document}